\documentclass[12pt]{iopart}

\expandafter\let\csname equation*\endcsname\relax
\expandafter\let\csname endequation*\endcsname\relax
\usepackage{graphics} 
\usepackage{graphicx}
\usepackage{dcolumn}
\usepackage{bm}
\usepackage{subfigure}
\usepackage{multirow}
\usepackage{float}
\usepackage{amsmath}
\usepackage{esvect}
\usepackage{autobreak}

\begin{document}

\title{Nuclear modification factor in Pb-Pb and p-Pb collisions using Boltzmann transport equation}
\author{Liyun Qiao, Guorong Che, Jinbiao Gu, Hua Zheng and Wenchao Zhang}
\address{School of Physics and Information Technology, Shaanxi Normal University, Xi'an 710119, People's Republic of China}
\eads{wenchao.zhang@snnu.edu.cn}
\begin{abstract}

  \noindent We investigate the nuclear modification factor ($R_{AA}$) of identified particles as a function of transverse momentum ($p_{\rm T}$)  in Pb-Pb collisions at $\sqrt{s_{\rm NN}}=$ 2.76 and 5.02 TeV, as well as p-Pb collision at 5.02 TeV in the framework of Boltzmann transport equation with relaxation time approximation. In this framework, the initial distribution of particles is chosen as the Tsallis distribution and the local equilibrium distribution  as the Boltzmann-Gibbs blast-wave distribution. The non-extensive parameter $q_{pp}$,  the Tsallis temperature $T_{pp}$ and the equilibrium temperature $T_{eq}$ are set to be in common for all particles, while the ratio of kinetic freeze-out time to relaxation time $t_{f}/\tau$ is different for different particles when we performed a combined fit to the $R_{AA}$ spectra of different particles at a given centrality. We observe that the fitted curves describe the spectra well up to $p_{\rm T}\approx$ 3 GeV/c. $q_{pp}$ and $T_{eq}$ ($t_{f}/\tau$) decrease (increases) with centrality nonlinearly, while $T_{pp}$ is almost independent of centrality. The dependence of the rate at which $q_{pp}$, $T_{eq}$ or $t_{f}/\tau$ changes with centrality on the energy and the size of the colliding system is discussed.

\end{abstract}

\maketitle

\section{Introduction}
\label{sec:intro}
Quantum-Chromodynamics (QCD) predicts that at high temperature and energy density there exists a hot and dense strongly interacting matter. This matter is commonly denoted as quark-gluon plasma (QGP), where partons (quarks and gluons) are the dominant degrees of freedom \cite{QGP}. QGP is expected to be produced in ultra-relativistic heavy-ion collisions. As the partons propagate through QGP, they lose energy due to gluon emission and parton splitting\footnote{In the paper, we focus on low momentum particles that originate from the QGP. Jet quenching effects are beyond the scope of this work.} \cite{energy_loss}. Understanding the mechanism of parton energy loss is thus one of the main goals in heavy-ion collisions. The energy loss can be investigated via the study of the difference of the transverse momentum ($p_{\rm T}$) spectrum in heavy-ion collisions compared to that in proton-proton (pp) collisions at the same energy. The difference is quantified by the nuclear modification factor $R_{AA(pA)}$, which is defined as \cite{RAA, RAA_1}
\begin{eqnarray}
R_{AA(pA)}(p_{\rm T})=\frac{1}{\langle {N_{\rm coll}} \rangle} \frac{(1/{2\pi p_{\rm T}}) d^{2} N_{AA(pA)}/dy dp_{\rm T}} {(1/{2\pi p_{\rm T}}) d^{2} N_{pp}/dy dp_{\rm T}},
\label{eq:RAA_exp}
\end{eqnarray}
where ${(1/{2\pi p_{\rm T}}) d^{2} N_{AA(pA)}/dy dp_{\rm T}}$ is the invariant $p_{\rm T}$ spectrum of nucleus-nucleus (AA) or proton-nucleus (pA) collisions, $(1/{2\pi p_{\rm T}})d^{2}N_{pp}/dy dp_{\rm T}$ is the invariant $p_{\rm T}$ spectrum in pp collisions. ${\langle {N_{\rm coll}} \rangle} $ is the average number of binary nucleon-nucleon collisions at a given centrality and is estimated by the Glauber model of the nuclear collision geometry \cite{Glauber}. In the absence of nuclear modification, $R_{AA(pA)}$ equals to unity. An observation of $R_{AA(pA)}$ deviating from 1 indicates the presence of in-medium effects.

As shown in refs. \cite{RAA_BTE, Sahho_1, Sahho_2}, the nuclear modification factor can also be expressed as 
\begin{eqnarray}
R_{AA(pA)}=\frac{f_{fin}}{f_{in}},
\label{eq:RAA_BTE}
\end{eqnarray}
where $f_{in}$ refers to the $p_{\rm T}$ distribution of particles produced immediately after collisions and  $f_{fin}$ refers to the final $p_{\rm T}$ distribution of particles. $f_{fin}$ can be obtained by plugging the initial distribution $f_{in}$ and the local equilibrium distribution $f_{eq}$ into Boltzmann transport equation (BTE) with relaxation time approximation (RTA) \cite{RTA}. In ref. \cite{Sahho_1}, $f_{in}$ was chosen as the Tsallis distribution \cite{Tsallis_dist} and $f_{eq}$ as the Boltzmann-Gibbs distribution. The expression in equation (\ref{eq:RAA_BTE}) then was fitted individually to the $R_{AA}$ spectra of pions, kaons, protons, $K_{S}^{0}$, $\rm \Lambda$, $D^0$ and $J/{\psi}$ (${\pi}^0$ and $D^0$) in the most central Pb-Pb (Au-Au) collisions at $\sqrt{s_{\rm NN}}=$ 2.76 TeV (200 GeV). For the heavy flavour hadrons, equation (\ref{eq:RAA_BTE}) can describe the $R_{AA}$ spectra well. However, for the light flavour hadrons,  it can only explain the $R_{AA}$ spectra in the intermediate to high $p_{\rm T}$ region. In ref. \cite{Sahho_2}, the authors took $f_{eq}$ as the Boltzmann-Gibbs blast-wave (BGBW) function  \cite{BGBW_dist}. They performed an individual fit to the $R_{AA}$ spectrum of pions, kaons, protons, $K^{*0}$ or ${\phi}$ in the most central Pb-Pb collisions at $\sqrt{s_{\rm NN}}=2.76$ TeV. It was observed that the $R_{AA}$ spectra of pions, kaons, $K^{*0}$ and ${\phi}$ can be well described, while not for protons.

In this paper, as a complementary study to that conducted in refs. \cite{Sahho_1, Sahho_2}, we investigate the $R_{AA}$ spectra of identified particles not only at the most central but also at the non-central Pb-Pb collisions at $\sqrt{s_{\rm NN}}=$ 2.76 TeV in the framework of BTE with  RTA. Similarly as done in ref. \cite{Sahho_2}, we set  $f_{in}$ as the Tsallis distribution and $f_{eq}$ as the BGBW distribution.  At a given centrality, we perform a combined rather than an individual fit on the $R_{AA}$ spectra of identified particles. In the combined fit, the following parameters are set to be in common for all particles: the non-extensive parameter $q_{pp}$, the Tsallis temperature $T_{pp}$, the equilibrium temperature $T_{eq}$, the average transverse velocity $\langle \beta \rangle$ and the exponent of the transverse velocity profile $n$\footnote{In the combined fit, $\langle \beta \rangle$ and $n$ are respectively fixed to values returned from a combined blast-wave fit on the pion, kaon and proton $p_{\rm T}$ spectra at the given centrality. See the explanation in section \ref{sec:results_and_discussions}.}.  Another parameter is the ratio of kinetic freeze-out time to relaxation time  $t_{f}/\tau$, which is different for different particles. We then extend our investigation to the $R_{AA}$ ($R_{\rm pPb}$) spectra of identified particles at different centralities in Pb-Pb (p-Pb) collisions at 5.02 TeV. The combined fit can provide insight on the degree of deviation from equilibrium for the system produced immediately after collisions, the temperature of the system at the local equilibrium and the time taken by the system to reach the equilibrium. Its usefulness lies in the ability to compare the results at different energies in the same colliding system and the results in different colliding systems at the same energy. From our study, we observe that the model can describe the $R_{AA}$ or $R_{\rm pPb}$ spectra of identified particles up to 3 GeV/c. Additionally, $T_{eq}$ and $q_{pp}$ ($t_{f}/\tau$) decrease (increases) with centrality, while $T_{pp}$  is almost independent of centrality.  The rate at which $T_{eq}$, $q_{pp}$ or $t_{f}/\tau$ changes with centrality depends on the energy and the size of the colliding system.

The organization of this paper is as follows. In section \ref{sec:method}, we briefly describe the derivation of the nuclear modification factor in the framework of BTE with RTA. In section \ref{sec:results_and_discussions}, we present the results of the combined fit to the $R_{AA}$ ($R_{\rm pPb}$) spectra of identified particles in Pb-Pb (p-Pb) collisions at 2.76 and 5.02 (5.02) TeV and make some discussions. Finally, the conclusion is given in section \ref{sec:conclusions}.

\section{Nuclear modification factor in the framework of BTE with RTA}\label{sec:method}

The derivation of the nuclear modification factor in the framework of BTE with RTA is described explicitly in refs. \cite{Sahho_1, Sahho_2}. Here we only show the result, 
\begin{eqnarray}
  R_{AA(pA)}=\frac{f_{eq}}{f_{in}}+\left(1-\frac{f_{eq}}{f_{in}}\right)e^{\frac{-t_f}{\tau}},
\label{eq:RAA_final_0}
\end{eqnarray}
where $t_{f}$ is the kinetic freeze-out time, $\tau $ is the relaxation time characterizing the time scale for the non-equilibrium system to relax to the local equilibrium. $f_{in}$ ($f_{eq}$, $f_{fin}$) is the particle distribution at $t=0$ ($\tau$, $t_{f}$) .

As described in ref. \cite{Tsallis_dist_1},  the system produced immediately after the high energy collisions usually stays away from thermal equilibrium. The temperature of the system $T_{B}$ fluctuates from event to event. Such a situation is described by a non-extensive statistics, i.e.,  the Tsallis statistics \cite{Tsallis_dist}. Therefore, as done in ref.  \cite{Sahho_2}, we set the initial distribution as the thermodynamically consistent  Tsallis distribution \cite{Tsallis_thermo}
\begin{eqnarray}
  f_{in}= C_{in}m_{\rm T}\left[1+(q_{pp}-1)\frac{p_{\rm T}}{T_{pp}}\right]^{-\frac{q_{pp}}{q_{pp}-1}},
  \label{eq:initial_dist}
\end{eqnarray}
where $m_{\rm T}$ is the transverse mass, $C_{in}=gV/(2 \pi)^2$, $g$ is the degeneracy factor, $V$ is the volume of the system. $T_{pp}$ is the Tsallis temperature, whose reciprocal represents the average value of $1/T_{B}$.  $q_{pp}$ is the non-extensive parameter, which is connected to the variance of $1/T_{B}$ \cite{Tsallis_dist_2}. It measures the degree of deviation from equilibrium.  The Tsallis distribution has been extensively used for the study of particle distributions in pp collisions \cite{Tsallis_thermo, Tsallis_thermo_1, Tsallis_thermo_2, Tsallis_thermo_3, Tsallis_thermo_4, Tsallis_thermo_5} and also in heavy ion collisions \cite{Tsallis_thermo_6, Tsallis_thermo_7, Tsallis_thermo_8, Tsallis_thermo_9, Tsallis_thermo_10}.

Once reaching the local equilibrium at the time $t=\tau$, the system will undergo the hydrodynamic evolution and finally freeze-out at the time $t=t_{f}$. Thus, as done in ref. \cite{Sahho_2}, we take the local equilibrium distribution as the BGBW function \cite{BGBW_dist}, 
\begin{eqnarray}
  f_{eq}=C_{eq}m_{\rm T}\int_{0}^{R_{0}}rdr K_{1}(\xi_{m})I_{0} (\xi_{p}),
   \label{eq:equlibrium_dist}
\end{eqnarray}
where $C_{eq}=2g \tau/(2\pi)^2$,  $K_{1}(\xi_{m})$ ($I_{0} (\xi_{p})$) is the modified Bessel function of the second (first) kind. $\xi_{m}=m_{\rm T} {\rm cosh}\rho/T_{eq}$, $\xi_{p}=p_{\rm T} {\rm sinh}\rho/T_{eq}$, $T_{eq}$ is the equilibrium temperature,  $\rho = {\rm tanh}^{-1}(\beta_{s}(r/R_{0})^n)$ is the transverse rapidity, $r$ is the radial distance, ${\beta}_s$ is the transverse flow velocity at the fireball surface ($r=R_{0}$), $n$ is the exponent of the velocity profile. The average transverse velocity is $\langle \beta \rangle=2/(n+2)\beta_s$. With the substitution of equations (\ref{eq:initial_dist}) and  (\ref{eq:equlibrium_dist}) into equation (\ref{eq:RAA_final_0}), the nuclear modification factor is written as
\begin{eqnarray}
  \begin{split}
\label{eq:RAA_final}
R_{AA(pA)}&=\frac{C_{eq}\int_{0}^{R_{0}}rdr K_{1}(\xi_{m})I_{0} (\xi_{p})}{C_{in} \left[1+(q_{pp}-1)\frac{p_{\rm T}}{T_{pp}}\right]^{-\frac{q_{pp}}{q_{pp}-1}}}\\
&+\left(1-\frac{C_{eq}\int_{0}^{R_{0}}rdr K_{1}(\xi_{m})I_{0} (\xi_{p})}{C_{in} \left[1+(q_{pp}-1)\frac{p_{\rm T}}{T_{pp}}\right]^{-\frac{q_{pp}}{q_{pp}-1}}}\right) e^{\frac{-t_f}{\tau}}.
\end{split}
\end{eqnarray}
The above expression incorporates a picture that particles produced in pp collisions undergo the evolution with a kinetic theory of BTE in RTA. 


\section{Results and discussions}\label{sec:results_and_discussions}
The ALICE collaboration have published the $R_{AA}$ spectra of pions, kaons and protons (the resonances $K^{*0}$ and $\phi$) at 0-5$\%$, 5-10$\%$, 10-20$\%$, 20-40$\%$, 40-60$\%$ and 60-80$\%$ (0-5$\%$, 5-10$\%$, 20-30$\%$ and 40-50$\%$) centralities in Pb-Pb collisions at 2.76 TeV in refs. \cite{pbpb_RAA,pbpb_RAA_1}. Here, the pion, kaon, proton and $K^{*0}$ spectra respectively refer to the spectra of $\pi^{+}+\pi^{-}$, $K^{+}+ K^{-}$, $p + \bar{p}$ and $K^{*0}+\bar{K}^{*0}$. For the $\rm \Lambda$ $R_{AA}$ spectra, no official data are released so far.  However, the ALICE collaboration have presented the $\rm \Lambda$ $p_{\rm T}$ spectra at 0-5$\%$, 5-10$\%$, 10-20$\%$, 20-40$\%$, 40-60$\%$ and 60-80$\%$ centralities in Pb-Pb collisions at this energy \cite{Lambda_pt_spectra}. Moreover, the preliminary result of the $\rm \Lambda$ $p_{\rm T}$ spectra in pp collisions at 2.76 TeV is available \cite{Lambda_pt_spectra_1}. Thus the $\rm \Lambda$ $R_{AA}$ spectra at these centralities can be constructed using equation (\ref{eq:RAA_exp}). The values of $\langle {N_{\rm coll}} \rangle$  in the denominator of this equation are taken from ref. \cite{pbpb_Ncoll_Npart}.   Recently, the ALICE collaboration have also published the $R_{AA}$ spectra of pions, kaons and protons at 0-5$\%$, 5-10$\%$, 10-20$\%$, 20-40$\%$, 40-60$\%$ and 60-80$\%$ centralities in Pb-Pb collisions at 5.02 TeV \cite{RAA_PbPb_5_02_identified}. The $R_{AA}$ spectra of $\rm \Xi$ and $\rm \Omega$ ($K^{*0}$, $\phi$, $\rm \Lambda$, $\rm \Xi$ and $\rm \Omega$) in Pb-Pb collisions at 2.76 (5.02) TeV are not considered in this work, as they are not available so far.

As the system size and number of particles produced in pA collisions are between those in pp and AA collisions, the results in pA collisions have frequently been utilized as a reference to understand those in AA collisions. In ref. \cite{p_Pb_Trans_moment}, the ALICE collaboration have published the $p_{\rm T}$ spectra of pions, kaons and protons at 0-5$\%$, 5-10$\%$, 10-20$\%$, 20-40$\%$, 40-60$\%$ and 60-80$\%$ centralities p-Pb collisions at 5.02 TeV. Moreover, they also presented the $p_{\rm T}$ spectra of these particles in pp collisions at this energy. Therefore, the $R_{\rm pPb}$ spectra at these centralities can be constructed using equation (\ref{eq:RAA_exp}). The $\langle N_{\rm coll} \rangle$ values are taken from ref. \cite{p_pb_Ncoll_Npart}. The ALICE collaboration have also published the $p_{\rm T}$ spectra of $K^{*0}$, $\phi$, $\rm \Lambda$, $\rm \Xi$ and $\rm \Omega$ at different centralities in p-Pb collisions at 5.02 TeV \cite{resonance_pPb, lambda_pPb, Xi_Omega_pPb}. However, the $p_{\rm T}$ spectra of these particles in pp collisions at this energy are not available so far. Therefore, in this work we do not consider their $R_{\rm pPb}$ spectra.

\begin{figure}[h]
  \centering
  \includegraphics[scale=0.195]{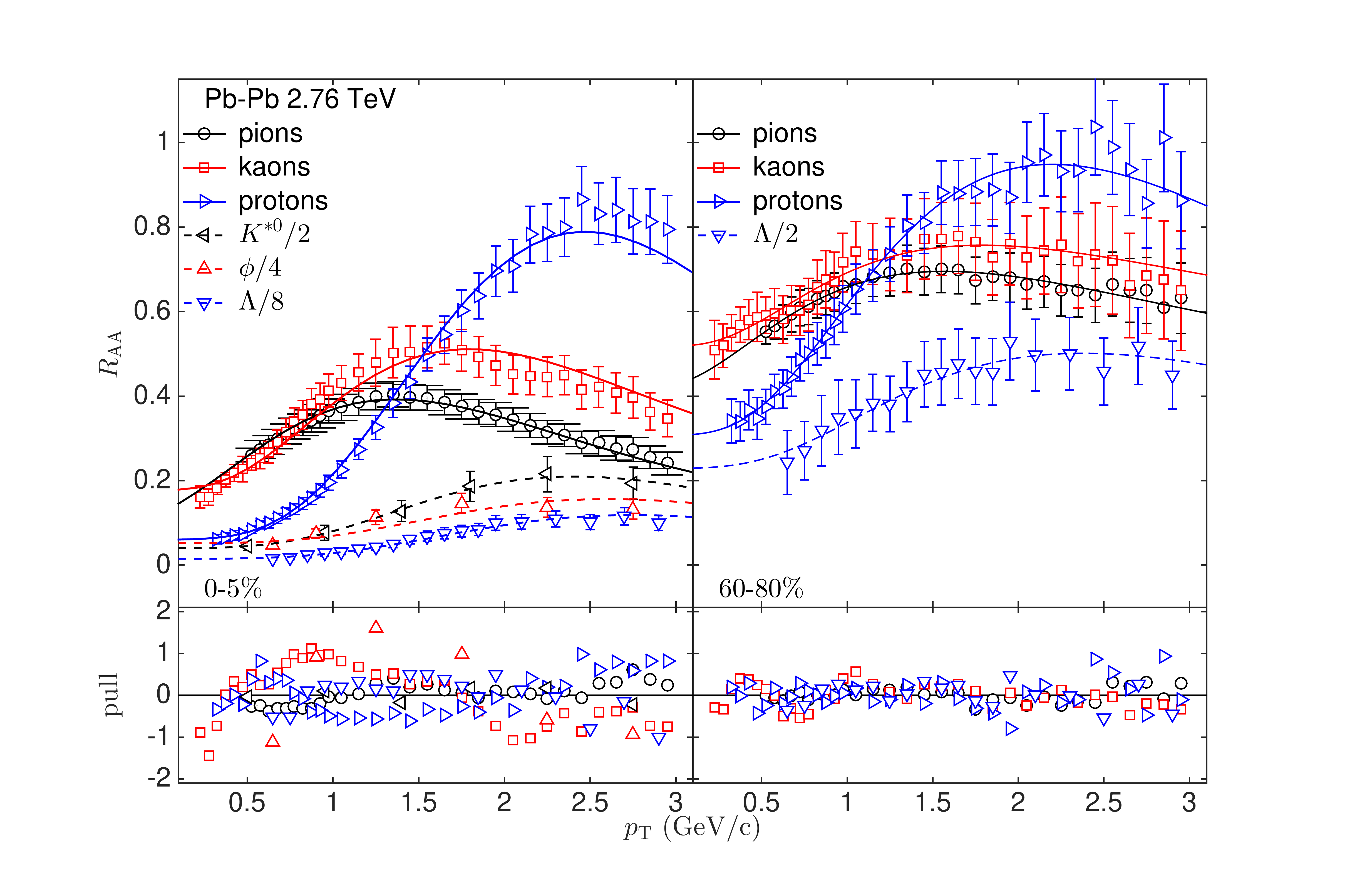}
 \caption{\label{fig:pb_pb_RAA_6_particle_plus_chi}(Colour online) Top left (right) panel: the $R_{AA}$ spectra of identified particles at the 0-5$\%$ (60-80$\%$) centrality in Pb-Pb collisions at 2.76 TeV. The data points are taken from refs. \cite{pbpb_RAA,pbpb_RAA_1,  Lambda_pt_spectra,Lambda_pt_spectra_1,blast_wave_pb_pb, pi_k_p_pp_2_76}. The curves represent the combined fit. Bottom left (right) panel: the pull distributions at the 0-5$\%$ (60-80$\%$) centrality.}
\end{figure}

We first perform a combined fit on the $R_{AA}$ spectra of pions, kaons, protons, $\rm \Lambda$, $K^{*0}$ and $\phi$ at the 0-5$\%$ centrality in Pb-Pb collisions at 2.76 TeV with equation (\ref{eq:RAA_final}) adopting a least $\chi^{2}$ method. At high $p_{\rm T}$, as a hard contribution may set in, the spectra are not expected to be described by the blast-wave model \cite{blast_wave_pb_pb}. Therefore, we limit the combined fit to $p_{\rm T}<$ 3 GeV/c\footnote{In ref. \cite{blast_wave_pb_pb}, the upper limit for pions (kaons, protons) is 1 (1.5, 3) GeV/c. The parameters returned from the combined fit with these upper limits are consistent with those in the paper within errors.}. At low $p_{\rm T}$, there is a large contribution from resonance decays for pions. In order to remove this contribution, we set the lower bound of the pion spectrum as 0.5 GeV/c, which is utilized by the experimental groups. In the fit, the statistical and systematic errors of the data points have been added in quadrature. As shown in ref. \cite{Tsallis_thermo}, the authors have fitted the $\pi^{\pm}$, $K^{\pm}$, $p(\bar{p})$, $K_{S}^{0}$, $\rm \Lambda$ and $\rm \Xi^{-}$ spectra produced in pp collisions at 0.9 TeV with the thermodynamically consistent Tsallis distribution individually. They found that for all hadrons the non-extensive parameter was around 1.11 and the Tsallis temperature was around 70 MeV. Thus, in the combined fit, $T_{pp}$ and  $q_{pp}$ are set to be in common for all particles. In ref. \cite{blast_wave_pb_pb}, in order to quantify the freeze-out parameters in Pb-Pb collisions at 2.76 TeV, a combined blast-wave fit of the pion, kaon and proton $p_{\rm T}$ spectra  was performed. Similarly as done in that reference, we set the parameters $T_{eq}$, $\langle \beta \rangle$ and $n$ for pions, kaons, protons, $\rm \Lambda$, $K^{*0}$ and $\phi$ in common in this work. As described in section \ref{sec:method}, the system will reach the local equilibrium at $t=\tau$, then it will undergo the hydrodynamic evolution and finally freeze out at $t=t_{f}$. It is reasonable to assume that  $\langle \beta \rangle$ and $n$ at $t=\tau$ are respectively the same as those at $t=t_{f}$. Therefore, we fix $\langle \beta \rangle$ and $n$ in $f_{eq}$ respectively as $0.651\pm0.020$ and $0.712\pm0.086$ which were returned by the combined blast-wave fit of the pion, kaon and proton $p_{\rm T}$ spectra at the 0-5$\%$ centrality \cite{blast_wave_pb_pb}. As the equilibration relies on the particle species and their interaction with the rest of the medium, the relaxation time differs from particles to particles \cite{Sahho_1}.  Thus we set $t_{f}/\tau$ to be different for different particles. As a result, there are nine parameters in the combined fit: three common parameters $q_{pp}$, $T_{pp}$ and $T_{eq}$; six parameters $t_{f}/\tau$, one for each particle. They are listed in table \ref{tab:pbpb_6_particles_fit_parameters}.  Also tabulated in the table is the $\chi^{2}$ per degree of freedom ($\chi^{2}/$dof)\footnote{The low $\chi^{2}/$dof probably indicates the experimental systematic errors are strongly correlated and thus overestimated.}. The first uncertainty quoted in the table is returned from the combined fit. The second is determined by adding the errors returned from the variation of $\langle \beta \rangle$ and $n$ in the combined fit respectively by $\pm 1\sigma$  in quadrature. The third is the uncertainty due to the variation of the lower fit bound (from 0.5 to 0.1 GeV/c) for pions. We then apply the same procedure to the $R_{AA}$ spectra of pions, kaons, protons, $\rm \Lambda$, $K^{*0}$ and $\phi$ at the 5-10$\%$ centrality. For the 10-20$\%$, 20-40$\%$, 40-60$\%$ or 60-80$\%$ centrality, we can only perform a combined fit on the  $R_{AA}$ spectra of pions, kaons, protons and $\rm \Lambda$, since the $R_{AA}$ spectra of $K^{*0}$ and $\phi$ at these centralities are not available so far. For the 20-30$\%$ or 40-50$\%$ centrality, we would like to perform a combined fit on the $R_{AA}$ spectra of pions, kaons, protons, $K^{*0}$ and $\phi$. At these two centralities, the $R_{AA}$ spectra of $K^{*0}$ and $\phi$  have been published, but not the $R_{AA}$ spectra of pions, kaons and protons.  However, we can construct their $R_{AA}$ spectra using equation (\ref{eq:RAA_exp}), since their $p_{\rm T}$ spectra in Pb-Pb and pp collisions at 2.76 TeV are available in refs.  \cite{blast_wave_pb_pb, pi_k_p_pp_2_76}. The parameters $q_{pp}$, $T_{pp}$, $T_{eq}$ and $t_{f}/\tau$ at these centralities are also tabulated in table \ref{tab:pbpb_6_particles_fit_parameters}. The upper panels of figure \ref{fig:pb_pb_RAA_6_particle_plus_chi} present the $R_{AA}$ spectra together with the combined fit results at two selected centralities (0-5$\%$ and 60-80$\%$). We observe that most of the data points appear consistent with the fitted curves which are described by equation (\ref{eq:RAA_final}). In order to address how much the combined fit is compatible with the data points statistically, a variable $\rm pull=\rm (data-fitted)/\Delta data$ is evaluated. The pull distributions are presented in the lower panels of the figure.  Except for the second point (the first and third points) in the  kaon ($\phi$) $R_{AA}$ spectrum at the 0-5$\%$ centrality, we observe that all the other data points are consistent with the fitted curves within one standard deviation. In the region with $p_{\rm T}>$ 3 GeV/c (not shown in the figure), there is a large deviation between the data points and the fitted curve for the $R_{AA}$ spectra of  kaons and protons.

\begin{figure}[h]
  \centering
  \includegraphics[scale=0.195]{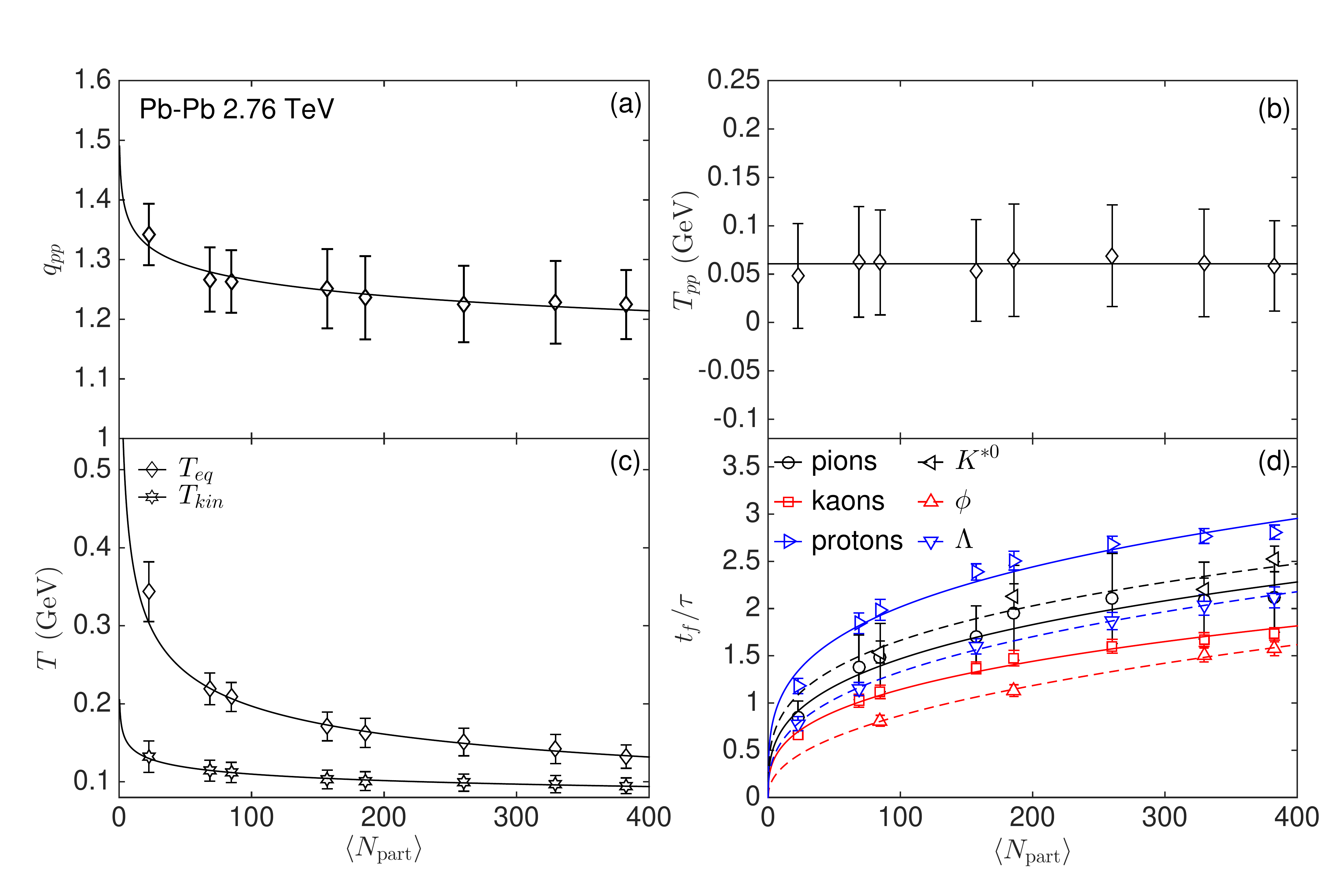}
 \caption{\label{fig:Pb_Pb_para_vs_Npart_merged}(Colour online) $q_{pp}$, $T_{pp}$,  $T_{eq}$, $T_{kin}$ and $t_{f}/\tau$ as a function of centrality in Pb-Pb collisions at 2.76 TeV. The error bar represents the total uncertainty of the parameter. Also shown in the figure is the parameterization of the dependence with $a\langle N_{\rm part} \rangle^{b}$. The values of $T_{kin}$ are taken from ref. \cite{blast_wave_pb_pb}.}
\end{figure}

With the parameters in table \ref{tab:pbpb_6_particles_fit_parameters}, we show $q_{pp}$, $T_{pp}$, $T_{eq}$ and $t_{f}/\tau$ versus centrality  in figure \ref{fig:Pb_Pb_para_vs_Npart_merged}. For $q_{pp}$, $T_{eq}$ and $t_{f}/\tau$, the dependence is clearly nonlinear and is parameterized with $a\langle N_{\rm part} \rangle^{b}$, where $a$ and $b$ are free parameters, $b$ represents the rate at which $ \textrm{ln} q_{pp}$, $\textrm{ln} T_{eq}$ or $ \textrm{ln} t_{f}/\tau$ changes with $\textrm{ln}\langle N_{\rm part} \rangle$. $\langle N_{\rm part} \rangle$ is the average value of the number of participants at a given centrality and is taken from ref. \cite{pbpb_Ncoll_Npart}. Several conclusions can be drawn from the results presented in the figure.

(i)  $q_{pp}$ decreases with  centrality. The $b$ value returned from the parameterization of $q_{pp}$ is $-0.030\pm0.005$. It means that the initial distribution in central collisions remains closer to equilibrium than that in peripheral collisions.

(ii)  $T_{pp}$ at different centralities are consistent within errors. They are around $60.6\pm 3.3$ MeV.

\begin{table}[H]
\footnotesize
\caption{\label{tab:pbpb_6_particles_fit_parameters}Values of parameters from the combined fit to the  $R_{AA}$ spectra of identified particles at different centralities in Pb-Pb collisions at 2.76 TeV. The uncertainties are explained in the text.}
\begin{center}
\begin{tabular}{ccc}
\hline
\textrm{\ }&
\textrm{$0-5$\%$$}&
\textrm{$5-10$\%$$}\\
\hline
\textrm{$q_{pp}$}& 1.225$\pm$0.007$\pm$0.057$\pm$0.004  & 1.228$\pm$0.008$\pm$0.069$\pm$0.004  \\
\textrm{$T_{pp}$}& 0.058$\pm$0.009$\pm$0.045$\pm$0.010 &0.062$\pm$0.010$\pm$0.054$\pm$0.009  \\
\textrm{$T_{eq}$}& 0.132$\pm$0.004$\pm$0.014$\pm$0.004 & 0.142$\pm$0.005$\pm$0.018$\pm$0.004  \\
\textrm{$(t_{f}/\tau)_{\pi}$}& 2.213$\pm$0.161$\pm$0.230$\pm$0.246 & 2.085$\pm$0.169$\pm$0.270$\pm$0.253  \\
\textrm{$(t_{f}/\tau)_{K}$}&1.731$\pm$0.033$\pm$0.051$\pm$0.023&1.672$\pm$0.032$\pm$0.063$\pm$0.022  \\
\textrm{$(t_{f}/\tau)_{p}$}&2.809$\pm$0.053$\pm$0.031$\pm$0.044&2.769$\pm$0.053$\pm$0.040$\pm$0.042  \\
\textrm{$(t_{f}/\tau)_{K^{*0}}$} & 2.526$\pm$0.134$\pm$0.019$\pm$0.016 & 2.199$\pm$0.121$\pm$0.017$\pm$0.009      \\
\textrm{$(t_{f}/\tau)_{\phi}$} & 1.574$\pm$0.074$\pm$0.001$\pm$0.008 & 1.508$\pm$0.073$\pm$0.001$\pm$0.010     \\
\textrm{$(t_{f}/\tau)_{\Lambda}$} & 2.118$\pm$0.106$\pm$0.024$\pm$0.026 &2.038$\pm$0.103$\pm$0.029$\pm$0.024 \\
\textrm{$\chi^{2}$/dof}&37.975/116&34.154/116\\

\hline
\textrm{\ }&
\textrm{$10-20$\%$$}&
\textrm{$20-30$\%$$}\\
\hline
\textrm{$q_{pp}$}& 1.226$\pm$0.007$\pm$0.064$\pm$0.005   &  1.236$\pm$0.009$\pm$0.068$\pm$0.014 \\
\textrm{$T_{pp}$}& 0.069$\pm$0.009$\pm$0.050$\pm$0.012  &  0.064$\pm$0.010$\pm$0.052$\pm$0.023\\
\textrm{$T_{eq}$}& 0.151$\pm$0.004$\pm$0.017$\pm$0.004 &  0.163$\pm$0.005$\pm$0.017$\pm$0.005  \\
\textrm{$(t_{f}/\tau)_{\pi}$}& 2.107$\pm$0.179$\pm$0.297$\pm$0.328 & 1.955$\pm$0.180$\pm$0.265$\pm$0.387 \\
\textrm{$(t_{f}/\tau)_{K}$}&1.601$\pm$0.028$\pm$0.062$\pm$0.029 & 1.476$\pm$0.029$\pm$0.060$\pm$0.048 \\
\textrm{$(t_{f}/\tau)_{p}$}&2.682$\pm$0.050$\pm$0.043$\pm$0.053\ & 2.507$\pm$0.050$\pm$0.041$\pm$0.076\\
\textrm{$(t_{f}/\tau)_{K^{*0}}$} & --- &  2.219$\pm$0.124$\pm$0.033$\pm$0.036    \\
\textrm{$(t_{f}/\tau)_{\phi}$} & ---&    1.131$\pm$0.061$\pm$0.004$\pm$0.012 \\
\textrm{$(t_{f}/\tau)_{\Lambda}$} & 1.868$\pm$0.085$\pm$0.021$\pm$0.026 & ---\\
\textrm{$\chi^{2}$/dof}& 22.329/108&17.608/99 \\

\hline
\textrm{\ }&
\textrm{$20-40$\%$$}&
\textrm{$40-50$\%$$}\\
\hline
\textrm{$q_{pp}$}& 1.251$\pm$0.007$\pm$0.065$\pm$0.011 & 1.263$\pm$0.012$\pm$0.044$\pm$0.026\\
\textrm{$T_{pp}$}& 0.054$\pm$0.009$\pm$0.048$\pm$0.018 & 0.062$\pm$0.013$\pm$0.038$\pm$0.037  \\
\textrm{$T_{eq}$}& 0.171$\pm$0.005$\pm$0.017$\pm$0.005& 0.209$\pm$0.008$\pm$0.016$\pm$0.004 \\
\textrm{$(t_{f}/\tau)_{\pi}$}& 1.706$\pm$0.108$\pm$0.179$\pm$0.245& 1.481$\pm$0.136$\pm$0.122$\pm$0.311 \\
\textrm{$(t_{f}/\tau)_{K}$}& 1.371$\pm$0.023$\pm$0.047$\pm$0.035 &1.117$\pm$0.027$\pm$0.036$\pm$0.055 \\
\textrm{$(t_{f}/\tau)_{p}$}& 2.387$\pm$0.043$\pm$0.041$\pm$0.063 & 1.986$\pm$0.050$\pm$0.041$\pm$0.090 \\
\textrm{$(t_{f}/\tau)_{K^{*0}}$} &--- & 1.540$\pm$0.107$\pm$0.026$\pm$0.036 \\
\textrm{$(t_{f}/\tau)_{\phi}$}& --- & 0.811$\pm$0.056$\pm$0.007$\pm$0.012 \\
\textrm{$(t_{f}/\tau)_{\Lambda}$} &1.599$\pm$0.072$\pm$0.022$\pm$0.026  & ---\\
\textrm{$\chi^{2}$/dof}&15.143/108  &13.719/99\\
\hline
 \textrm{\ }&
 \textrm{$40-60$\%$$}& 
 \textrm{$60-80$\%$$} \\
 \hline
 \textrm{$q_{pp}$}   &1.267$\pm$0.010$\pm$0.045$\pm$0.028  & 1.342$\pm$0.026$\pm$0.039$\pm$0.022\\
 \textrm{$T_{pp}$}   & 0.063$\pm$0.011$\pm$0.040$\pm$0.039  & 0.048$\pm$0.024$\pm$0.035$\pm$0.034 \\ 
 \textrm{$T_{eq}$}   & 0.219$\pm$0.008$\pm$0.018$\pm$0.004  & 0.344$\pm$0.028$\pm$0.025$\pm$0.003  \\
 \textrm{$(t_{f}/\tau)_{\pi}$} &  1.378$\pm$0.109$\pm$0.122$\pm$0.302  &  0.851$\pm$0.073$\pm$0.041$\pm$0.149  \\
 \textrm{$(t_{f}/\tau)_{K}$}   & 1.023$\pm$0.022$\pm$0.036$\pm$0.054   &  0.658$\pm$0.021$\pm$0.016$\pm$0.026  \\
 \textrm{$(t_{f}/\tau)_{p}$}  & 1.843$\pm$0.042$\pm$0.045$\pm$0.091     &  1.180$\pm$0.044$\pm$0.034$\pm$0.058\\ 
 \textrm{$(t_{f}/\tau)_{\Lambda}$}&1.149$\pm$0.062$\pm$0.011$\pm$0.027 &0.780$\pm$0.066$\pm$0.017$\pm$0.033\\
 \textrm{$\chi^{2}$/dof}  &10.241/108 & 9.090/108  \\
 \hline
\end{tabular}
\end{center}
\end{table}

(iii) $T_{eq}$ decreases with centrality. This trend is similar to that of the kinetic freeze-out temperature $T_{kin}$ returned from the combined blast-wave fit on the pion, kaon and proton $p_{\rm T}$ spectra in Pb-Pb collisions at 2.76 TeV \cite{blast_wave_pb_pb}. At a given centrality,  $T_{eq}$ is larger than $T_{kin}$, which is consistent with the picture that the temperature decreases with the evolution of the system from the local equilibrium to the kinetic freeze-out. Moreover, the rate at which $\textrm{ln} T_{eq}$ decreases with $\textrm{ln}\langle N_{\rm part} \rangle$ is $0.290\pm0.012$, which is larger than that of $\textrm{ln} T_{kin}$, $0.113\pm0.007$.

(iv) $t_{f}/\tau$ increases with centrality. It means that the relaxation time $\tau$ in central collisions is less than that in peripheral collisions. As the initial distribution in central collisions is less off-equilibrium, the system will accordingly take less time to reach the local equilibrium. The $b$ values for pions, kaons, protons, $K^{*0}$, $\phi$ and $\rm \Lambda$ are $0.318\pm0.035$, $0.334\pm0.022$, $0.275\pm0.033$, $0.287\pm0.170$, $0.450\pm0.044$ and $0.356\pm0.012$ respectively. Moreover, at a given centrality, we observe that for mesons (baryons or resonances) $t_{f}/\tau$ is smaller for particles with heavier mass, which is in agreement with the conclusion in ref. \cite{Sahho_1}.

\begin{figure}[h]
  \centering
  \includegraphics[scale=0.195]{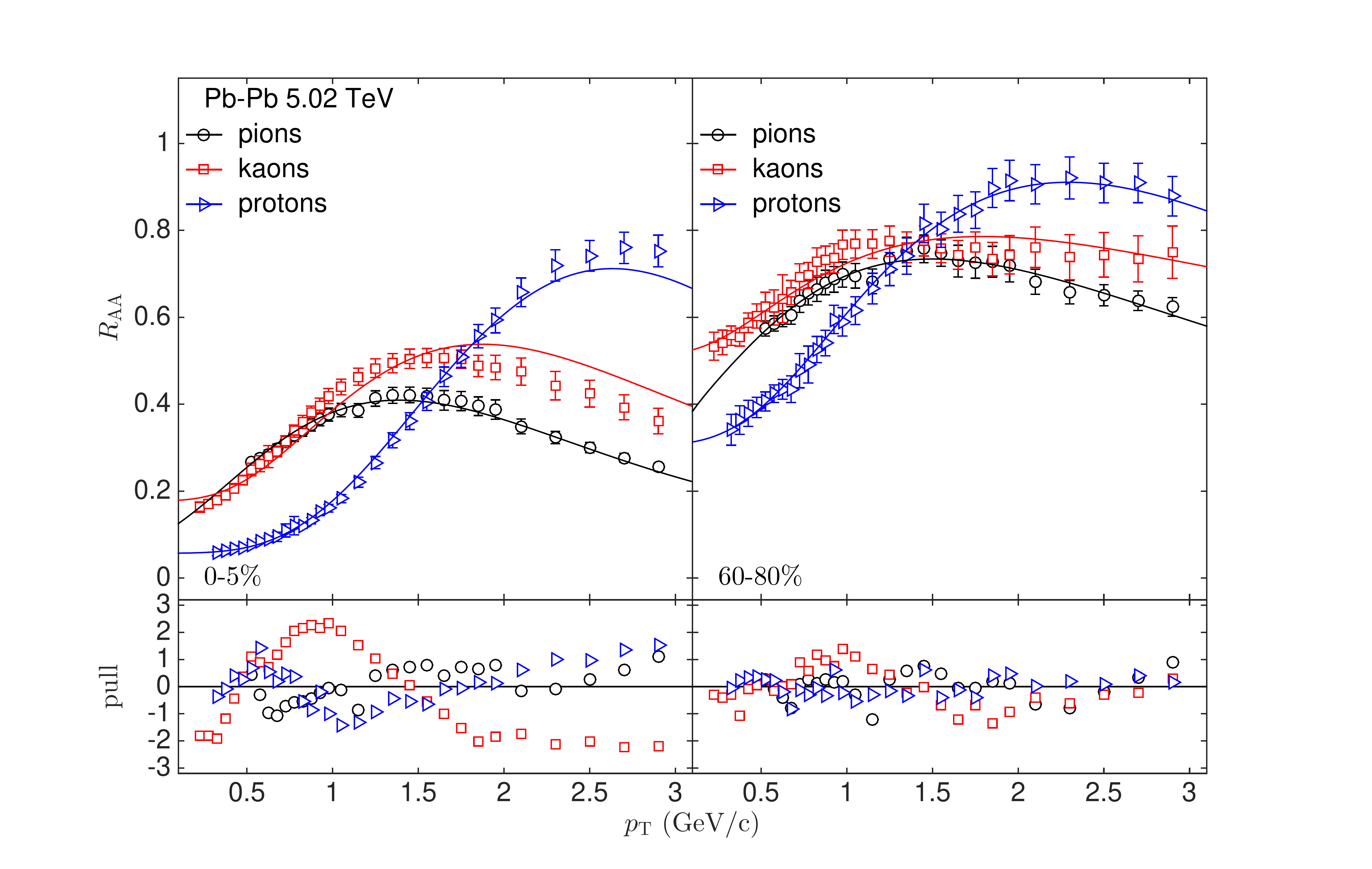}
 \caption{\label{fig:pb_pb_RAA_3_particle_plus_chi} (Colour online) Top left (right) panel: the $R_{AA}$ spectra of pions, kaons and protons at the 0-5$\%$ (60-80$\%$) centrality in Pb-Pb collisions at 5.02 TeV. The data points are taken from ref. \cite{RAA_PbPb_5_02_identified}. The curves represent the combined fit. Bottom left (right) panel: the pull distributions at the 0-5$\%$ (60-80$\%$) centrality.}
\end{figure}

\begin{figure}[t]
  \centering
  \includegraphics[scale=0.195]{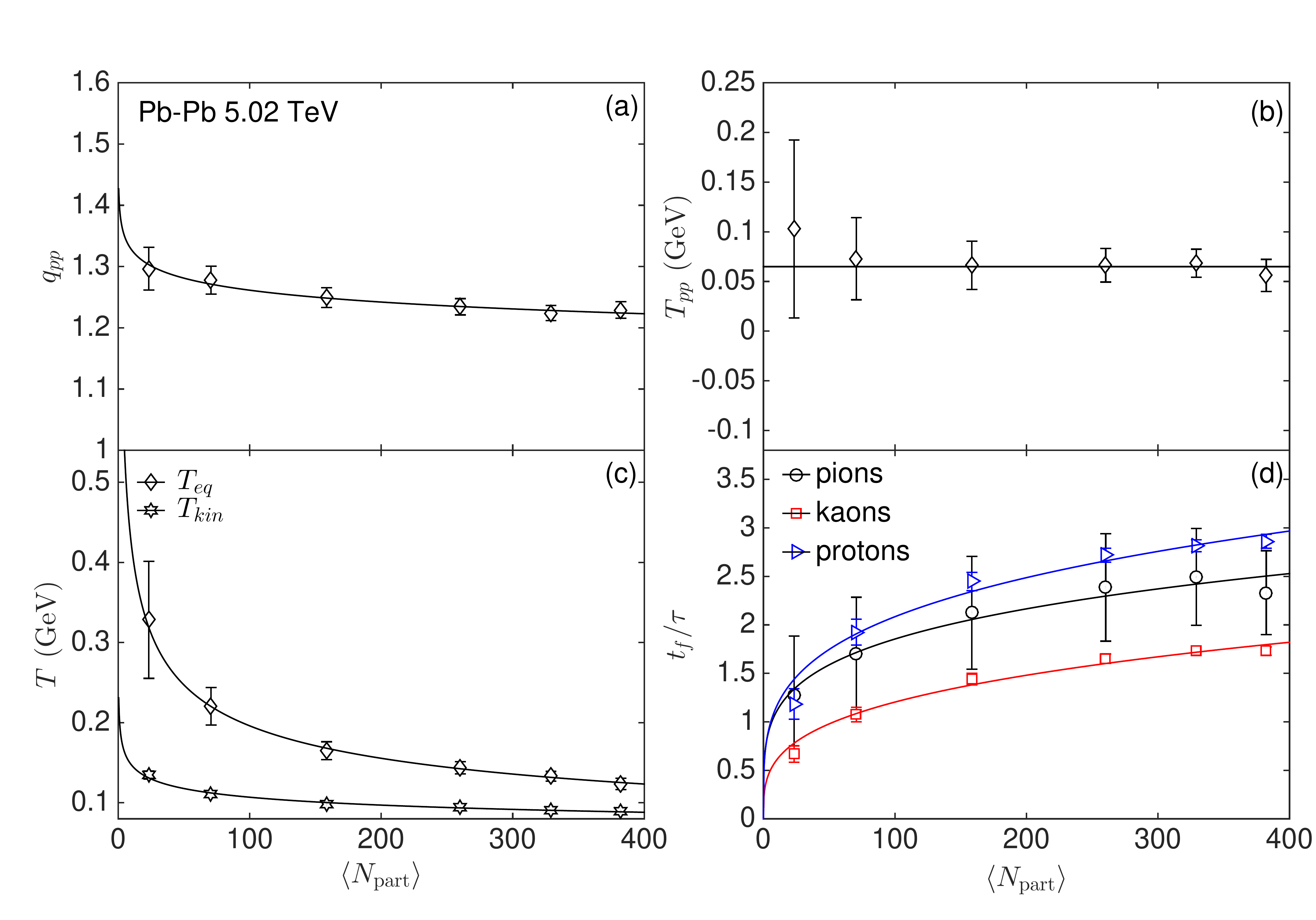}
 \caption{\label{fig:Pb_Pb_para_vs_Npart_merged_5_02_TeV}(Colour online) $q_{pp}$, $T_{pp}$,  $T_{eq}$, $T_{kin}$ and $t_{f}/\tau$ as a function of centrality  in Pb-Pb collisions at 5.02 TeV. The error bar represents the total uncertainty of the parameter. Also shown in the figure is the parameterization of the dependence with $a\langle N_{\rm part} \rangle^{b}$. The values of $T_{kin}$ are taken from ref. \cite{RAA_PbPb_5_02_identified}.}
\end{figure}

Next we investigate the $R_{AA}$ spectra of pions, kaons and protons at the 0-5$\%$ centrality in Pb-Pb collisions at 5.02 TeV. We fix $\langle \beta \rangle$ and $n$ respectively as $0.663\pm0.003$ and $0.735\pm0.013$ which were returned by a combined blast-wave fit of the pion, kaon and proton $p_{\rm T}$ spectra at this centrality in ref. \cite{RAA_PbPb_5_02_identified}. Combined fits on the $R_{AA}$ spectra at other centralities are performed similarly as that at the 0-5$\%$ centrality. In the upper panels of figure \ref{fig:pb_pb_RAA_3_particle_plus_chi}, we present the $R_{AA}$ spectra together with the fit results at two selected centralities (0-5$\%$ and 60-80$\%$).  As can be seen from the pull distributions in the lower panels of the figure, at the 60-80$\%$ (0-5$\%$) centrality most of the data points agree with the fitted curves within 1 (2) standard deviation(s). The parameters returned from the combined fits and their uncertainties are reported in table \ref{tab:pb_pb_3_particles_fit_parameters}. Also listed in the table is the $\chi^{2}/$dof\footnote{The $\chi^{2}/$dof increases with centrality, probably indicating the experimental uncertainties in peripheral collisions are more overestimated than in central collisions.}. The dependence of these parameters on centrality is shown in figure \ref{fig:Pb_Pb_para_vs_Npart_merged_5_02_TeV}. In the figure, the values of $\langle N_{\rm part} \rangle$ at different centralities are taken from ref. \cite{Pb_Pb_5_02_Npart}. As can be seen from the figure, $q_{pp}$, $T_{eq}$, $T_{kin}$ and $t_{f}/\tau$ rely nonlinearly  on centrality. This trend is similar to that observed in Pb-Pb collisions at 2.76 TeV. Moreover,  at a given centrality, we observe that $q_{pp}$, $T_{eq}$, $T_{kin}$, $t_{f}/\tau$ for pions, kaons and protons are respectively compatible with those at 2.76 TeV within errors. However, the rate at which $ \textrm{ln} q_{pp}$ ($\textrm{ln} T_{eq}$, $\textrm{ln} T_{kin}$) changes with $\textrm{ln}\langle N_{\rm part} \rangle$ is $-0.022\pm 0.004$ ($-0.334\pm 0.018$, $-0.140\pm 0.014$), whose absolute value is smaller (larger, larger) than that at 2.76 TeV. This means at high energy $q_{pp}$ ($T_{eq}$, $T_{kin}$) decreases slower (faster, faster) with centrality than that at low energy in the same colliding system. For pions (kaons, protons), the rate at which $ \textrm{ln} t_{f}/\tau$ increases with $\textrm{ln}\langle N_{\rm part} \rangle$  is $0.224\pm 0.052$ ($0.298\pm 0.056$, $0.255\pm 0.049$), which is smaller than (compatible with, compatible with) that at 2.76 TeV. For $T_{pp}$, it fluctuates around $64.8\pm 3.7$ MeV, which is consistent with that at 2.76 TeV within errors. The conclusion that at a given centrality $t_{f}/\tau$ is smaller for particles with heavier mass also holds for mesons (pions and kaons) in Pb-Pb collisions at 5.02 TeV.

\begin{table}[t]
  \caption{\label{tab:pb_pb_3_particles_fit_parameters}Values of parameters from the combined fit to the pion, kaon and proton $R_{AA}$ spectra at different centralities in Pb-Pb collisions at 5.02 TeV. $T_{pp}$ and $T_{eq}$ are in units of GeV. The uncertainties are the same as those in table \ref{tab:pbpb_6_particles_fit_parameters}.}
\footnotesize
\begin{center}
  \begin{tabular}{cccc}
\hline
\textrm{\ }&
\textrm{$0-5$\%$$}&
\textrm{$5-10$\%$$}&
\textrm{$10-20$\%$$}\\
\hline
\textrm{$q_{pp}$}& 1.229$\pm$0.011$\pm$0.008$\pm$0.001&1.2241$\pm$0.0095$\pm$0.0077$\pm$0.0002&1.234$\pm$0.011$\pm$0.007$\pm$0.002  \\
\textrm{$T_{pp}$}& 0.056$\pm$0.014$\pm$0.006$\pm$0.006&0.068$\pm$0.013$\pm$0.006$\pm$0.003& 0.066$\pm$0.014$\pm$0.005$\pm$0.008 \\
\textrm{$T_{eq}$}& 0.124$\pm$0.006$\pm$0.002$\pm$0.004 & 0.133$\pm$0.005$\pm$0.002$\pm$0.003& 0.144$\pm$0.006$\pm$0.002$\pm$0.004\\ 
\textrm{$(t_{f}/\tau)_{\pi}$}& 2.332$\pm$0.313$\pm$0.051$\pm$0.295 & 2.496$\pm$0.390$\pm$0.061$\pm$0.305 &  2.385$\pm$0.390$\pm$0.051$\pm$0.389\\  
\textrm{$(t_{f}/\tau)_{K}$}&1.729$\pm$0.040$\pm$0.007$\pm$0.014&1.731$\pm$0.040$\pm$0.007$\pm$0.010& 1.652$\pm$0.043$\pm$0.007$\pm$0.022\\   
\textrm{$(t_{f}/\tau)_{p}$}&2.8623$\pm$0.0663$\pm$0.0005$\pm$0.0272&2.8156$\pm$0.0578$\pm$0.0005$\pm$0.0183 & 2.718$\pm$0.063$\pm$0.001$\pm$0.036\\  
\textrm{$\chi^{2}$/dof}&110.408/76&76.928/76&86.283/76\\
 \hline
 \textrm{\ }&
 \textrm{$20-40$\%$$}&
 \textrm{$40-60$\%$$}& 
 \textrm{$60-80$\%$$} \\
 \hline
 \textrm{$q_{pp}$}      & 1.249$\pm$0.012$\pm$0.007$\pm$0.008 &1.278$\pm$0.015$\pm$0.008$\pm$0.015      & 1.297$\pm$0.019$\pm$0.011$\pm$0.027\\
 \textrm{$T_{pp}$}  & 0.066$\pm$0.015$\pm$0.005$\pm$0.018 & 0.073$\pm$0.018$\pm$0.006$\pm$0.037     & 0.103$\pm$0.026$\pm$0.009$\pm$0.085 \\ 
 \textrm{$T_{eq}$}   & 0.165$\pm$0.007$\pm$0.002$\pm$0.008& 0.220$\pm$0.013$\pm$0.003$\pm$0.019    & 0.328$\pm$0.035$\pm$0.009$\pm$0.064 \\
  \textrm{$(t_{f}/\tau)_{\pi}$}    &  2.124$\pm$0.348$\pm$0.037$\pm$0.464 &  1.704$\pm$0.289$\pm$0.029$\pm$0.503      &  1.274$\pm$0.262$\pm$0.027$\pm$0.549  \\
  \textrm{$(t_{f}/\tau)_{K}$}   &  1.441$\pm$0.043$\pm$0.006$\pm$0.042& 1.075$\pm$0.040$\pm$0.006$\pm$0.064  &  0.667$\pm$0.033$\pm$0.006$\pm$0.078  \\
  \textrm{$(t_{f}/\tau)_{p}$} &  2.448$\pm$0.067$\pm$0.002$\pm$0.066 & 1.925$\pm$0.072$\pm$0.006$\pm$0.113   &  1.185$\pm$0.063$\pm$0.010$\pm$0.144\\ 
  \textrm{$\chi^{2}$/dof} & 104.950/76&68.931/76& 24.874/76
  \\
  \hline
\end{tabular}
\end{center}
\end{table}

\begin{figure}[h]
  \centering
  \includegraphics[scale=0.195]{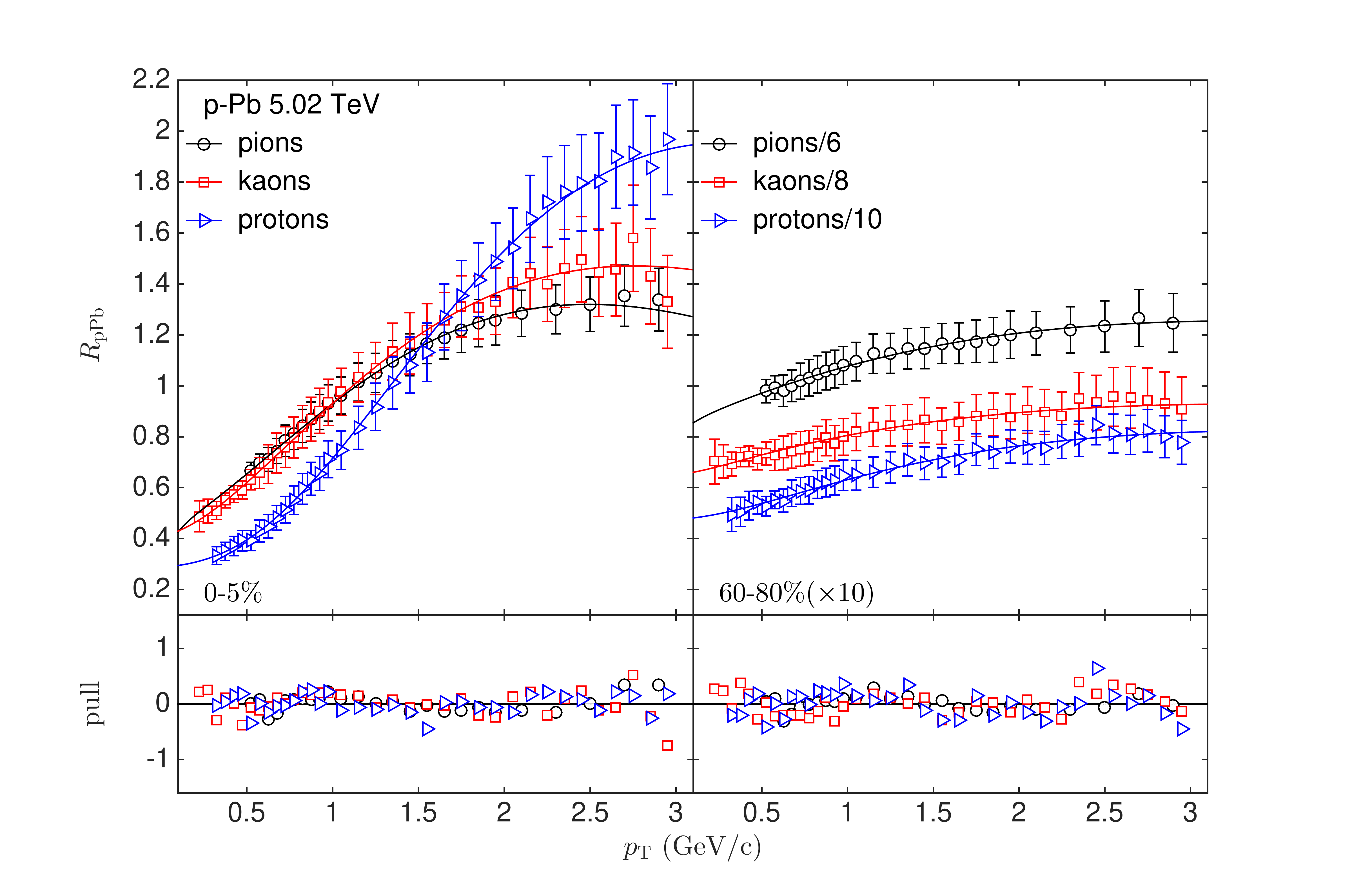}
 \caption{\label{fig:p_pb_RAA_3_particle_plus_chi}(Colour online) Top left (right) panel: the $R_{\rm pPb}$ spectra of pions, kaons and protons at the 0-5$\%$ (60-80$\%$) centrality in p-Pb collisions at 5.02 TeV. The data points are taken from ref. \cite{p_Pb_Trans_moment}. The curves represent the  combined fit. Bottom left (right) panel: the pull distributions at the 0-5$\%$ (60-80$\%$) centrality.}
\end{figure}

\begin{figure}[t]
   \centering
   \includegraphics[scale=0.195]{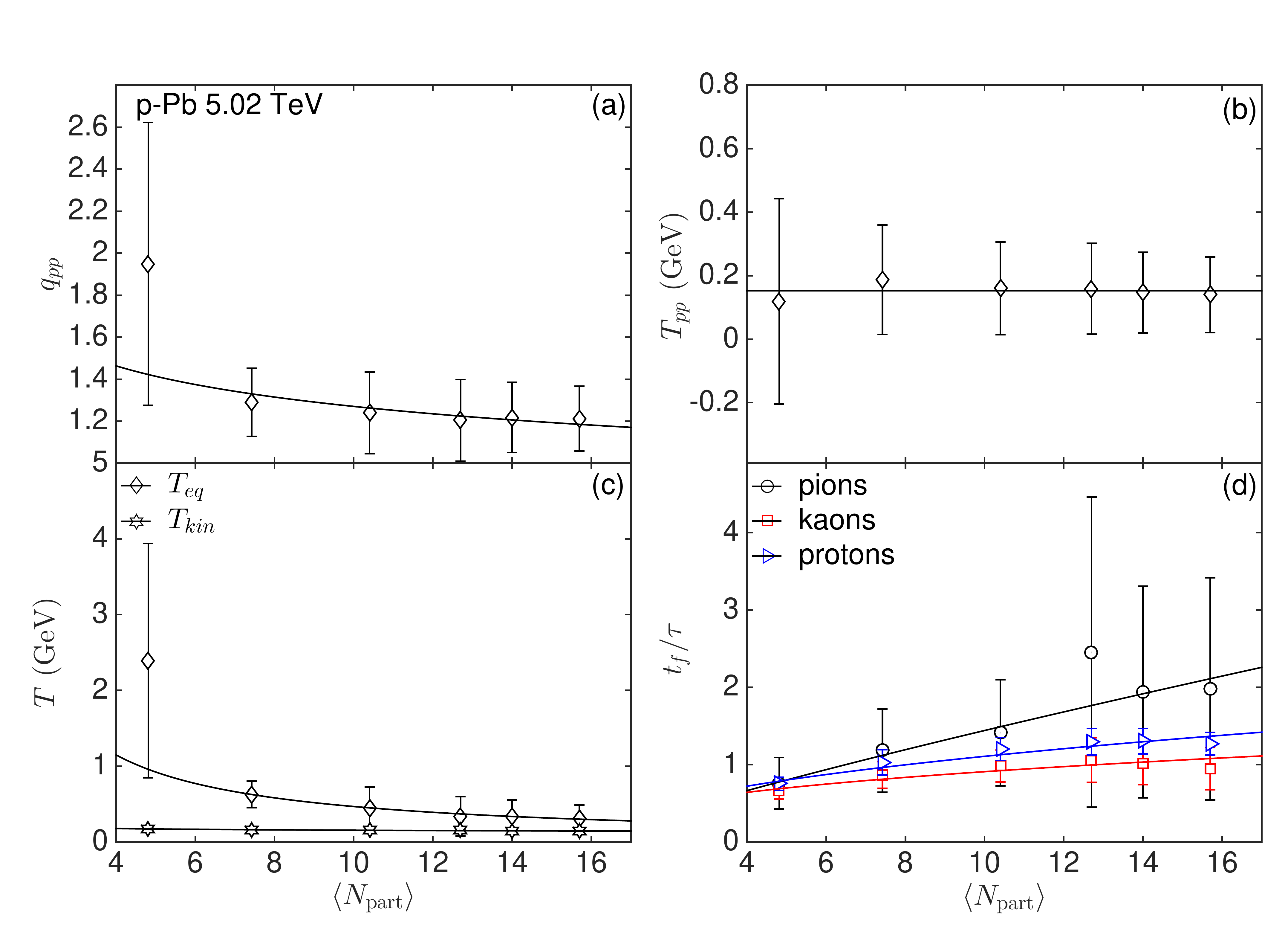}
 \caption{\label{fig:p_Pb_para_vs_Npart_merged}(Colour online) $q_{pp}$, $T_{pp}$,  $T_{eq}$, $T_{kin}$ and $t_{f}/\tau$ as a function of centrality  in p-Pb collisions at 5.02 TeV. The error bar represents the total uncertainty of the parameter. Also shown in the figure is the parameterization of the dependence with $a\langle N_{\rm part} \rangle^{b}$. The values of $T_{kin}$ are taken from ref. \cite{lambda_pPb}.}
\end{figure}

Finally, we extend the investigation to the $R_{\rm pPb}$ spectra of pions, kaons and protons at a given centrality in p-Pb collisions at 5.02 TeV. In the combined fit, $\langle \beta \rangle$ and $n$ are fixed as the corresponding values that were returned by a combined blast-wave fit of the pion, kaon, proton, $K_{S}^{0}$ and $\rm \Lambda$ $p_{\rm T}$ spectra at this centrality in ref. \cite{lambda_pPb}. The combined fit results on the $R_{\rm pPb}$ at two selected centralities (0-5$\%$ and 60-80$\%$) are presented in the upper panels of figure \ref{fig:p_pb_RAA_3_particle_plus_chi}. From the lower panels of the figure, we observe that the data points are compatible with the fitted curves within 1 standard deviation at both centralities. The parameters returned from the combined fit and their uncertainties are listed in table \ref{tab:p_pb_3_particles_fit_parameters}. The dependence of these parameters on centrality is shown in figure \ref{fig:p_Pb_para_vs_Npart_merged}. In the figure, the values of $\langle N_{\rm part} \rangle$ at different centralities are taken from ref. \cite{p_pb_Ncoll_Npart}. The error bars of $q_{pp}$, $T_{pp}$ and $T_{eq}$ at the 60-80$\%$ centrality are relatively large. The main contributions to these error bars are the uncertainties returned from the fit and from the variation of the lower fit bound for pions. As can be seen from the figure, the dependence of  $q_{pp}$, $T_{eq}$, $T_{kin}$ and $t_{f}/\tau$ on centrality is nonlinear, which is similar to that in Pb-Pb collisions at 5.02 TeV. However, the rate at which $ \textrm{ln} q_{pp}$ ($\textrm{ln} T_{eq}$, $\textrm{ln} T_{kin}$) changes with $\textrm{ln}\langle N_{\rm part} \rangle$ is $-0.154\pm 0.086$ ($-0.980\pm 0.283$, $-0.148\pm 0.025$), whose absolute value is larger than (larger than, compatible with) that in Pb-Pb collisions at 5.02 TeV. Thus, in the small system $q_{pp}$ and $T_{eq}$ ($T_{kin}$) decrease(s) with centrality faster than (at the similar rate as) those (that) in the large system at the same energy.  For pions (kaons, protons), the rate at which $ \textrm{ln} t_{f}/\tau$ increases with $\textrm{ln}\langle N_{\rm part} \rangle$  is $0.846\pm 0.160$ ($0.380\pm 0.120$, $0.468\pm 0.100$), which is larger than (compatible with, larger than) that in Pb-Pb collisions at 5.02 TeV.  For $T_{pp}$, it fluctuates around $152.8\pm 10.4$ MeV, which is larger than that in Pb-Pb collisions at 5.02 TeV. Moreover, for mesons, at a given centrality $t_{f}/\tau$ is smaller for heavy particles (kaons) than that for light particles (pions), which is consistent with the conclusions drawn in Pb-Pb collisions at 2.76 and 5.02 TeV.



\begin{table}[t]
  \caption{\label{tab:p_pb_3_particles_fit_parameters} Values of parameters from the combined fit to the pion, kaon and proton $R_{\rm pPb}$ spectra at different centralities in p-Pb collisions at 5.02 TeV. $T_{pp}$ and $T_{eq}$ are in units of GeV. The uncertainties  are the same as those in table \ref{tab:pbpb_6_particles_fit_parameters}.
}
\footnotesize
\begin{center}
   \begin{tabular}{cccc}
\hline
\textrm{\ }&
\textrm{$0-5$\%$$}&
\textrm{$5-10$\%$$}&
\textrm{$10-20$\%$$}\\
\hline
\textrm{$q_{pp}$}& 1.213$\pm$0.011$\pm$0.021$\pm$0.153  & 1.218$\pm$0.014$\pm$0.036$\pm$0.163  & 1.204$\pm$0.013$\pm$0.033$\pm$0.191  \\
\textrm{$T_{pp}$}& 0.140$\pm$0.008$\pm$0.013$\pm$0.118 &0.146$\pm$0.010$\pm$0.027$\pm$0.124& 0.159$\pm$0.009$\pm$0.024$\pm$0.141 \\
\textrm{$T_{eq}$}& 0.309$\pm$0.015$\pm$0.009$\pm$0.177 & 0.341$\pm$0.022$\pm$0.018$\pm$0.210 & 0.337$\pm$0.023$\pm$0.020$\pm$0.257\\ 

\textrm{$(t_{f}/\tau)_{\pi}$}& 1.980$\pm$0.502$\pm$0.159$\pm$1.336 & 1.937$\pm$0.506$\pm$0.358$\pm$1.220 &  2.454$\pm$0.935$\pm$0.518$\pm$1.697\\  
\textrm{$(t_{f}/\tau)_{K}$}&0.948$\pm$0.041$\pm$0.030$\pm$0.267&1.011$\pm$0.047$\pm$0.063$\pm$0.259 & 1.060$\pm$0.046$\pm$0.056$\pm$0.280\\  
 
\textrm{$(t_{f}/\tau)_{p}$}&1.271$\pm$0.024$\pm$0.014$\pm$0.144&1.303$\pm$0.030$\pm$0.031$\pm$0.159 & 1.295$\pm$0.028$\pm$0.026$\pm$0.167\\ 
\textrm{$\chi^{2}$/dof}&3.281/86&4.585/86&4.280/86\\
 \hline
 \textrm{\ }&
 \textrm{$20-40$\%$$}&
 \textrm{$40-60$\%$$}& 
 \textrm{$60-80$\%$$} \\
 \hline
 \textrm{$q_{pp}$}  & 1.240$\pm$0.019$\pm$0.030$\pm$0.191 &1.290$\pm$0.060$\pm$0.028$\pm$0.148  & 1.949$\pm$0.509$\pm$0.018$\pm$0.441\\
 \textrm{$T_{pp}$}  & 0.160$\pm$0.011$\pm$0.020$\pm$0.144 & 0.188$\pm$0.026$\pm$0.017$\pm$0.170  & 0.119$\pm$0.250$\pm$0.009$\pm$0.205 \\ 
 \textrm{$T_{eq}$}   & 0.448$\pm$0.040$\pm$0.022$\pm$0.272& 0.628$\pm$0.127$\pm$0.043$\pm$0.113  & 2.392$\pm$1.214$\pm$0.062$\pm$0.957 \\
 
  \textrm{$(t_{f}/\tau)_{\pi}$} & 1.411$\pm$0.186$\pm$0.101$\pm$0.652  &  1.181$\pm$0.223$\pm$0.051$\pm$0.487  &  0.760$\pm$0.138$\pm$0.006$\pm$0.303  \\
  \textrm{$(t_{f}/\tau)_{K}$}  & 0.982$\pm$0.034$\pm$0.031$\pm$0.200 & 0.866$\pm$0.049$\pm$0.017$\pm$0.166   &  0.661$\pm$0.060$\pm$0.002$\pm$0.087 \\
  \textrm{$(t_{f}/\tau)_{p}$} & 1.203$\pm$0.025$\pm$0.019$\pm$0.147 & 1.029$\pm$0.036$\pm$0.014$\pm$0.158     &  0.752$\pm$0.057$\pm$0.002$\pm$0.065\\ 
  
  \textrm{$\chi^{2}$/dof} & 2.880/86 &4.178/86 & 3.717/86  \\
 
  \hline
\end{tabular}
\end{center}
\end{table}

\section{Conclusions}
\label{sec:conclusions}
In this paper, we have investigated the $R_{AA}$ ($R_{\rm pPb}$) spectra of identified particles at different centralities in Pb-Pb (p-Pb) collisions at 2.76  and 5.02 (5.02) TeV in the framework of BTE with RTA. In this framework, $f_{in}$ is set to be the Tsallis distribution and $f_{eq}$ to be the BGBW distribution. At a given centrality, a combined fit is performed on the spectra of identified particles with a least $\chi^{2}$ method. In the combined fit, $q_{pp}$,  $T_{pp}$ and  $T_{eq}$ are set to be in common for all particles, while $t_{f}/\tau$ is different for different particles. We observe the fitted curves can describe the $R_{AA}$ or $R_{\rm pPb}$ spectra well up to $p_{\rm T} \approx$ 3 GeV/c. $q_{pp}$ and $T_{eq}$ ($t_{f}/\tau$) decrease (increases) with centrality nonlinearly. In the same colliding system, at high energy $q_{pp}$ ($T_{eq}$) decreases  slower (faster), while $t_{f}/\tau$ does not increase faster with centrality than that at low energy. At the same energy, in the large system $q_{pp}$ ($T_{eq}$) decrease slower (slower), while $t_{f}/\tau$ does not increase faster with centrality than that in the small system. $T_{pp}$ is almost independent of centrality, fluctuating around  $60.6\pm 3.3$ and $64.8\pm 3.7$ ($152.8\pm 10.4$) MeV in Pb-Pb (p-Pb) collisions at 2.76 and 5.02 (5.02) TeV.

\section*{Acknowledgements}
This work is supported by the Fundamental Research Funds for the Central Universities of China under GK201903022 and GK202003019, by the Scientific Research Foundation for the Returned Overseas Chinese Scholars, State Education Ministry, by Natural Science Basic Research Plan in Shaanxi Province of China (program No. 2020JM-289) and by the National Natural Science Foundation of China under Grant Nos. 11447024 and 11505108.


\end{document}